\documentclass{emulateapj}
\usepackage{graphicx}

\shorttitle{Evolutionary model for super-Chandrasekhar white dwarfs}
\shortauthors{Das, Mukhopadhyay \& Rao}

\begin{document}

\title{A possible evolutionary scenario of highly magnetized super-Chandrasekhar white dwarfs:
progenitors of peculiar type~Ia supernovae}

\author{Upasana Das\altaffilmark{1}, Banibrata Mukhopadhyay\altaffilmark{1}, A. R. Rao\altaffilmark{2}}
\altaffiltext{1}{Department of Physics, Indian Institute of Science, Bangalore 560012, India; 
upasana@physics.iisc.ernet.in, bm@physics.iisc.ernet.in} 
\altaffiltext{2}{Department of Astronomy \& Astrophysics, Tata Institute of Fundamental Research, 
Mumbai  400005, India; arrao@tifr.res.in}

\begin{abstract}
Several recently discovered peculiar type~Ia supernovae seem to
demand an altogether new formation theory that might help explain
the puzzling dissimilarities between them and the standard type~Ia supernovae.
The most striking aspect of the observational analysis is the
necessity of invoking super-Chandrasekhar white
dwarfs having masses $\sim 2.1-2.8M_{\odot}$, $M_\odot$ being the mass of Sun, 
as their most probable progenitors. 
Strongly magnetized white dwarfs having super-Chandrasekhar masses were already established to be 
potential candidates for the progenitors of peculiar type~Ia supernovae. Owing to the Landau quantization of the 
underlying electron degenerate gas, theoretical results yielded the observationally inferred mass range.
Here we sketch a possible evolutionary scenario by which  super-Chandrasekhar white dwarfs 
could be formed by accretion on to a commonly observed magnetized white dwarf, 
invoking the phenomenon of flux freezing. This opens the multiple possible evolutions ending in supernova explosions of super-Chandrasekhar white dwarfs having
masses within the range stated above. We point out that our proposal has
observational support, like, the recent discovery of a large number of magnetized white dwarfs by SDSS.

\end{abstract}


\section{Introduction}

Identifying the progenitors of type~Ia supernovae is
an extremely important and ongoing research issue \cite{howell2011}. 
These supernovae are considered to be standard candles for cosmic distance 
measurements and hence are probes for studying the expansion history of the 
universe \cite{perl99}. According to the general consensus, 
a carbon-oxygen white dwarf in a binary system accretes matter from a companion 
star and gradually approaches the Chandrasekhar mass limit of $1.44M_{\odot}$ \cite{chandra35},
$M_{\odot}$ being the mass of Sun. 
This further triggers a thermonuclear instability in the white dwarf, resulting in a violent 
and extremely energetic explosion, which we observe as a type~Ia supernova. However, 
far less is known about the nature of the companion star \cite{partha}. 

Adding to the puzzle further is the recent discovery of several peculiar type~Ia supernovae --- 
SN 2006gz, SN 2007if, SN 2009dc, SN 2003fg \cite{scalzo}. These 
supernovae are distinctly overluminous, powered by a higher than usual production 
of nickel, and have very low ejecta velocity compared to their standard counterparts \cite{nature}. 
They also violate the luminosity-stretch relation \cite{phillips,gold} which prohibits them 
from being categorized as standard candles. 
Interestingly, however, these anomalies seem to be resolved 
if instead of the standard theory one invokes super-Chandrasekhar-mass white dwarfs, 
with masses $2.1-2.8M_{\odot}$, as their progenitors \cite{scalzo,nature,hicken,yam,silverman,
taub}. The leading 
question now is the origin and stability of these super-Chandrasekhar white dwarfs. 
There have been a few simulations of accreting binary white dwarfs, which include
differential rotation and other parameters \cite{kato}, that try to explain the 
required mass range stated above. In addition, models have been proposed involving 
mergers of two white dwarfs, namely the double degenerate scenario \cite{tutu,scalzo},
and core degenerate scenario \cite{kashi}. However, a detailed theoretical understanding seems to be lacking
in them.

On a fundamentally different ground, Das \& Mukhopadhyay (2012a, 2012b) have shown
that strongly magnetized white dwarfs, having central fields $\sim 10^{15}-10^{17}$ G, 
are capable of having super-Chandrasekhar masses in the range $2-2.6M_{\odot}$. 
The basic idea employed (Das \& Mukhopadhyay 2012a, 2012b, hereafter DM1, DM2 respectively) is that, 
for magnetic fields ($B$) greater 
than a critical field $B_c= 4.414 \times 10^{13}$ G, the effect of Landau quantization 
on the underlying
electron degenerate matter becomes significant and 
as a result the density of states 
for the electrons changes \cite{lai}. This in turn modifies the corresponding 
equation of state (EoS) and hence the mass-radius relation of these white dwarfs. 
Now the additional question is to form white dwarfs with such strong magnetic fields.

In this {\it Letter}, we sketch an evolutionary scenario by which highly magnetized
white dwarfs could be formed. 
We concentrate on the single degenerate progenitor scenario.
Starting with some reasonable assumptions about the mass and the magnetic field of
a normally observed magnetic white dwarf, we demonstrate that by the mechanism of flux freezing and 
Landau quantization a possible evolutionary link between strongly magnetized 
{\it supermassive} white dwarfs and the observed peculiar type~Ia supernovae
could be established. This scenario is compatible with 
the discovery of several (isolated) magnetized white 
dwarfs by SDSS having surface fields $10^5-10^9$ G \cite{schmidt03,vanlandingham05} 
and the fact that $25\%$ of the observed cataclysmic variables (CVs)
have surface field strengths as high as $10^7-10^8$ G \cite{wick}. 

This {\it Letter} is organized as follows. In \S 2 we describe how a magnetized, 
sub-Chandrasekhar, accreting white dwarf evolves into a super-Chandrasekhar white dwarf. 
We subsequently discuss the steps that lead 
to the formation of multiple evolutionary tracks in the mass-radius plane, which in turn 
help in explaining the observed super-Chandrasekhar (progenitor) mass range of the peculiar 
type~Ia supernovae. In \S 3 we study the timescale of evolution of 
these magnetized accreting white dwarfs.
Finally we conclude in \S 4 with a discussion about the observational evidence 
supporting our proposed evolutionary scenario.

\section{Evolutionary path from a sub-Chandrasekhar to a super-Chandrasekhar white dwarf}

\begin{figure}[h]
\vskip 0.3in
   \centering
\includegraphics[scale=0.26]{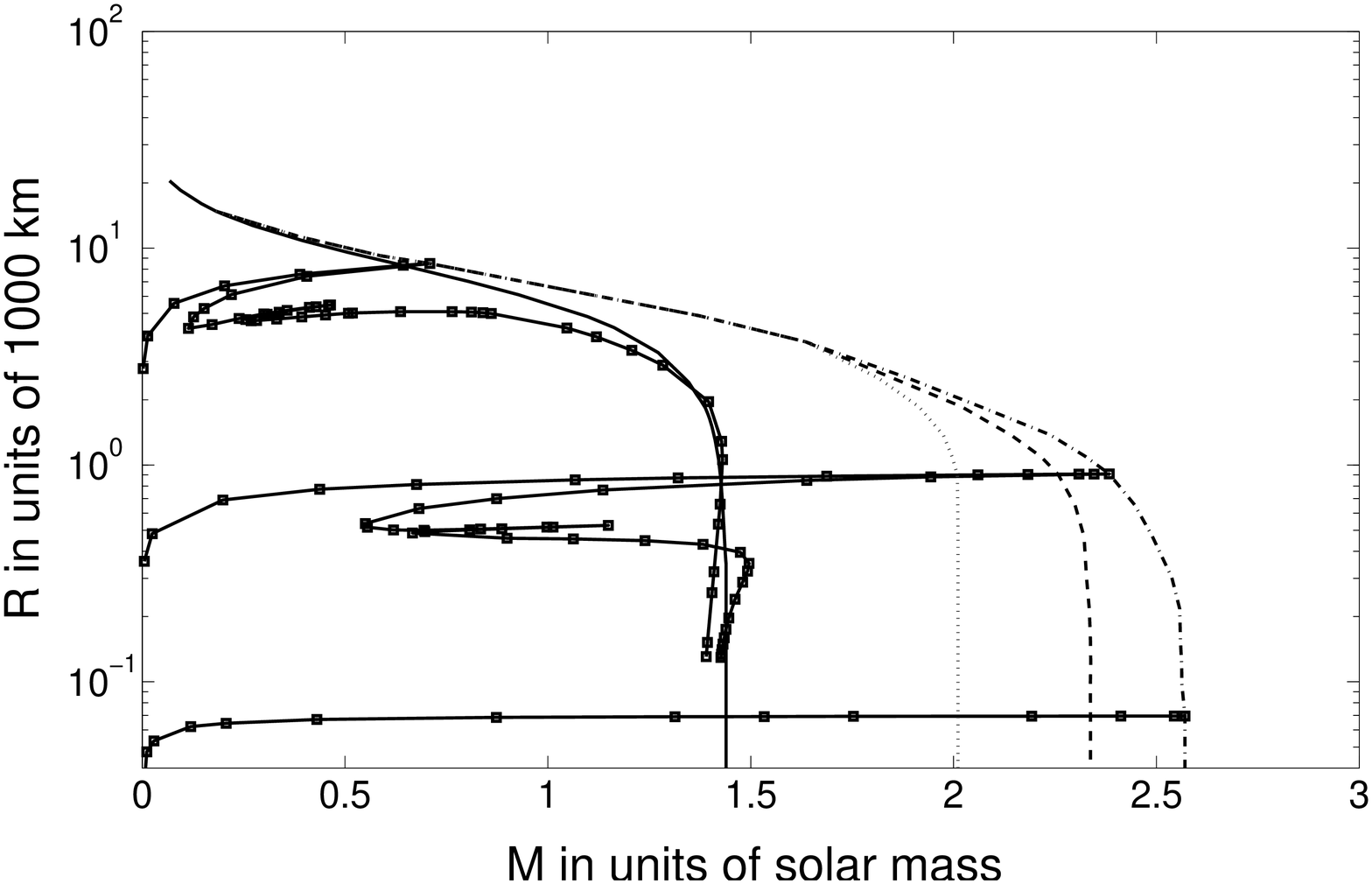}
\vskip0.5cm
\caption{Mass-Radius relations --- solid line
represents Chandrasekhar's non-magnetic result; dotted, dashed and
dot-dashed lines represent the evolutionary 
tracks of accreting magnetized white dwarfs, ending in super-Chandrasekhar type~Ia 
supernovae with progenitor masses $2.01 M_{\odot}$, $2.33 M_{\odot}$ and $2.58 M_{\odot}$ respectively. 
The solid lines marked with squares represent from top to bottom the intermediate 
mass-radius relations corresponding to 50124-level, 200-level and 1-level systems of Landau quantization respectively.}
\label{mr}
\end{figure}

We consider a white dwarf having a central field several orders of magnitude higher 
than that of the surface, with a typical mass and radius
to be $M_{0}\sim 0.2 M_{\odot}$ and $R \sim 15000$ km respectively. 
Note, as justified in \S IV.C of DM1, that the magnetic
field might be approximately constant in a range of radii around the center of the white dwarf, which we define
as internal magnetic field.  
This is quite a plausible assumption, since the original star 
collapsing to form a white dwarf might have a very large interior magnetic field
compared to that observed on its surface, as has been the case for the Sun 
\cite{gough}. Thus, it is quite likely that the values
of the central and surface magnetic fluxes (in the initial star and hence in the
white dwarf) would be significantly different, developing a core-envelope boundary.
Hence, we start with an internal magnetic field $B_{int}\sim5\times 10^{12}$~G along with a surface field $B_s$ at least three orders of magnitude lower than $B_{int}$. 
Such a white dwarf still 
has $B<B_c$ and  it lies on Chandrasekhar's (non-magnetic) mass-radius relation having EoS similar 
to that of a nonmagnetic case. 
Now as the white dwarf accretes matter, there is an amplification of its 
central magnetic field as a consequence of the increase in 
central density (via flux freezing theorem) due to the contraction 
in size of the white dwarf \cite{cumming}. This leads the
central magnetic field to eventually exceed $B_{c}$. Hence, EoS 
of the underlying electron degenerate gas and the mass-radius relation of the 
underlying white dwarf modify, as shown in DM1 and DM2. As a result, the white dwarf transits from the 
mass-radius curve for the nonmagnetic EoS to that for a magnetic EoS. As the  outward
modified pressure counteracts the inward gravitational force, 
a quasi-equilibrium state attains, which is determined by 
the degree of Landau quantization of the system. Larger the magnetic field, 
smaller is the number of Landau levels occupied by the electrons, leading to 
more massive white dwarfs (DM1, DM2). Hence, as accretion continues, 
the mass-radius curve of the white dwarf deviates 
more and more from that of Chandrasekhar's. In Figure \ref{mr}, we show three mass-radius relations 
(solid curves marked with squares) following DM1, for magnetized white dwarfs
with three different values of $B_{int}$ --- $1.7\times 10^{13}$~G, $4.4\times 10^{15}$~G and $8.8\times 10^{17}$~G,
for the top to bottom curves respectively, through which the evolving white dwarf, 
with initial mass, radius and magnetic fields mentioned above, passes at different intermediate stages. 

Following Bandyopadhyay et al. (1997), we adopt a profile for magnetic field
in our study assuming the surface and central magnetic fluxes of the initial and final
white dwarfs are individually conserved as the evolution progresses.
For example, for the white dwarf with $B_{int}=8.8\times 10^{17}$~G
and $R=69.5$~km (which is the maximum possible radius for the corresponding mass-radius relation), if we consider 
$B_{int}$ to be confined upto $R_{int}=R/10$ 
(corresponding to a profile with parameters $\beta=0.01,\gamma=3.6$ \cite{bando}),
then the above assumption holds true 
with surface and central fluxes
respectively $1.46\times 10^{27}$~G cm$^2$ and $4.25\times 10^{29}$~G cm$^2$. Note, 
however, that the extent of $B_{int}$ adjusts
accordingly as the white dwarf evolves and hence $R_{int}$ need not be the same for the initial and 
final white dwarfs.
Figure \ref{mr} shows how the an initial mass of the 
white dwarf could possibly evolve to $3$ different final limiting masses, namely 
$2.01M_{\odot}$, $2.33M_{\odot}$ and $2.58M_{\odot}$. 
Once the corresponding final mass is reached, further accretion induces a runaway 
thermonuclear reaction, causing the white dwarf to explode at that mass, 
leading to a {\it super-Chandrasekhar supernova}. 
We note that, although we show only three tracks in Figure \ref{mr}, any number of 
tracks are possible resulting in different final limiting 
masses $\lesssim 2.58M_{\odot}$. 
In the following subsection we explain in detail how we obtain these multiple tracks.

\subsection{Multiple evolutionary tracks}

When the electrons occupy the ground Landau level, pressure increases 
monotonically as a function of density. This corresponds to a mass-radius relation 
where initially the mass increases with increasing radius 
(let us call it the first branch of the mass-radius relation). 
If the magnetic field is sufficiently high, then the radius becomes nearly independent 
of mass at higher central densities in this branch. 
From DM1, we note that whenever there is a transition from a lower Landau level to the next higher level, 
a kink appears in EoS, followed by a plateau --- which is a small 
region where the pressure becomes nearly independent of the density (see Figure 1 of DM1). 
For a given magnetic field, the central density of the white dwarf corresponding to the kink in EoS 
arising at the transition from the ground to first Landau levels 
corresponds to a white dwarf having the maximum possible radius. 
Note that the systems having more than one 
occupied Landau level exhibit multiple branches in their mass-radius relations 
including the `first branch' (for details see DM1). 
For example, the mass-radius relation of white dwarfs having central densities 
corresponding to the plateau of EoS shows a turning point and subsequent 
decrease of mass with decreasing radius (see Figure 2 of DM1).

Now, let us consider a white dwarf lying on an intermediate mass-radius relation 
in its early stage of evolution (e.g. the top most solid curve marked with squares in Figure 
\ref{mr}), such that it has the maximum possible radius. 
As it evolves via accretion, its magnetic field and mass increases with decreasing radius, such 
that when the white dwarf 
transits to the next mass-radius relation, it 
still has the corresponding maximum radius satisfying the flux-freezing condition. 
For a sufficiently high field, the white dwarf with maximum radius also corresponds to the maximum possible 
mass.
Thus, in Figure 1, we show that the evolution of such a white dwarf 
gives rise to the track corresponding to the limiting mass of $2.58M_{\odot}$, which 
corresponds to the new mass limit for white dwarfs proposed by 
Das \& Mukhopadhyay (2013). Hence, this track demarcates a zone in the mass-radius plane, 
beyond the right-hand side of which no further track is possible.

From the bottom two mass-radius relations (representing further intermediate masses and radii
of the evolving white dwarf) in Figure 1,
we notice that the evolving white dwarf passing through them  may
lie anywhere in a zone where the radius is independent of mass satisfying the flux freezing
condition. This zone is a part of 
the `first branch' of the corresponding mass-radius relation.
This happens at a sufficiently high field $\gtrsim 1.7 \times 10^{14}$ G (which corresponds to a 
$5000$-Landau-level system), when a
part of EoS can be described by a polytropic relation of the form
\begin{equation}
P=K \rho^{\Gamma},~ {\rm when~~ \Gamma =2},
\end{equation}
where $P$ and $\rho$ are respectively the 
pressure and density of the electron gas, $\Gamma$ is the polytropic index 
and $K$ is a constant depending on the magnetic field. Now on 
solving the magnetostatic equilibrium condition by Lane-Emden 
formalism, one obtains the following scaling laws (DM1) for radius ($R$) and mass ($M$):
\begin{equation}
 R \propto \rho_{c}^{(\Gamma-2)/2}
\end{equation}
and
\begin{equation}
M \propto \rho_{c}^{(3\Gamma-4)/2}, 
\end{equation}
where $\rho_{c}$ is the central density of the white dwarf. 
Thus when $\Gamma=2$, the radius becomes independent of the central density, while 
the mass becomes proportional to the central density. This indicates 
that the same magnetic field leads to a white dwarf having different 
possible masses with the same radius. Hence once the above polytropic
relation starts holding true, the flux freezing condition opens
the multiple possible evolutionary tracks.

\section{Timescale of mass evolution}

Here we explore the typical timescale of evolution of the accreting magnetic white dwarf systems 
with assumed parameters 
quite close to the class of magnetic CVs called the Intermediate Polars
(IPs).

\subsection{Constant mass accretion rate}
\begin{figure}[h]
\vskip 0.3in
   \centering
\includegraphics[scale=0.26]{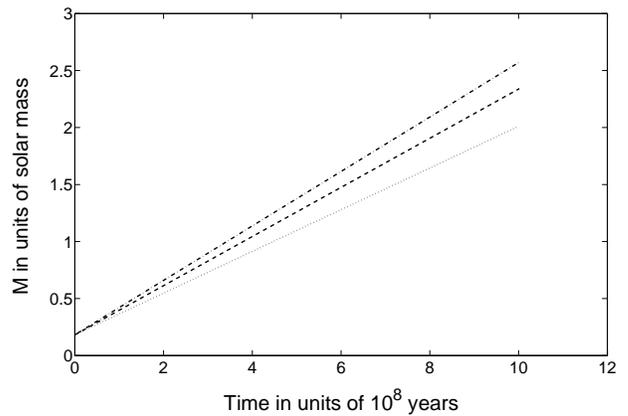}
\vskip0.5cm
\caption{Mass evolution of accreting magnetized white dwarfs --- 
dot-dashed, dashed and dotted lines represent constant accretion rates 
of respectively $2.38 \times 10^{-9} M_{\odot} {\rm yr^{-1}}$, $2.15 \times 10^{-9} M_{\odot} {\rm yr^{-1}}$ and 
$1.83 \times 10^{-9} M_{\odot} {\rm yr^{-1}}$ for supernova explosions occurring at 
the same time, for progenitor masses $2.58 M_{\odot}$, $2.33 M_{\odot}$ 
and $2.01 M_{\odot}$ respectively.}
\label{mtsame}
\end{figure}

Figure \ref{mtsame} shows the evolution of an initial white dwarf of mass $0.2 M_{\odot}$ 
and radius $15000$ km into the 3 super-Chandrasekhar white dwarfs with $M\ge2M_\odot$ 
corresponding to Figure \ref{mr}. Here we assume 
a constant mass accretion rate ($\dot{M}$), represented by the constant slopes of the 3 lines in Figure \ref{mtsame},
for each case, such that they explode at the same time. 
The values of $\dot{M}$
(typical of IPs) are $2.38 \times 10^{-9} M_{\odot} {\rm yr^{-1}}$, 
$2.15 \times 10^{-9} M_{\odot} {\rm yr^{-1}}$ and 
$1.83 \times 10^{-9} M_{\odot} {\rm yr^{-1}}$ for the exploding masses of $2.58 M_{\odot}$, $2.33 M_{\odot}$ 
and $2.01 M_{\odot}$ respectively. We see that the system with a higher $\dot{M}$ accumulating more mass 
in the same time (roughly a billion year) gives rise to a more massive supernova explosion.

Figure 3(a) shows that how the exploding mass of $2.33 M_{\odot}$ is attained by 3 different $\dot{M}$s
 --- $3 \times 10^{-9} M_{\odot} {\rm yr^{-1}}$, $2.5 \times 10^{-9} M_{\odot} {\rm yr^{-1}}$ 
and $2 \times 10^{-9} M_{\odot} {\rm yr^{-1}}$ --- 
resulting in the explosions occurring at different times.

\subsection{Varying mass accretion rate}

\begin{figure}[h]
\vskip 0.3in
   \centering
\includegraphics[scale=0.26]{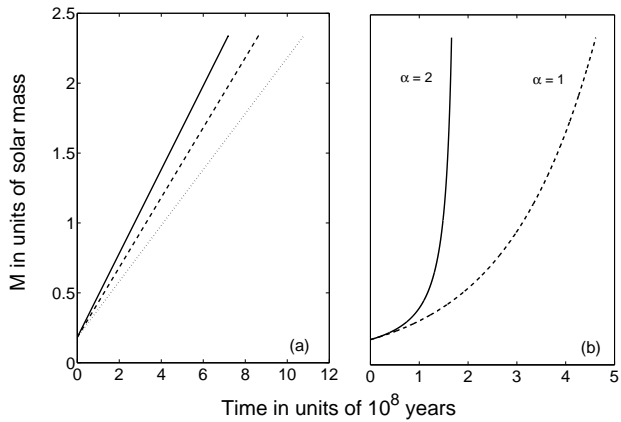}
\vskip0.5cm
\caption{Mass evolution of accreting magnetized white dwarf with (a) constant accretion rate --- solid, dashed and dotted lines represent $3 \times 10^{-9} M_{\odot} {\rm yr^{-1}}$, $2.5 \times 10^{-9} M_{\odot} {\rm yr^{-1}}$ and $2 \times 10^{-9} M_{\odot} {\rm yr^{-1}}$ respectively, (b) varying accretion rate --- solid and dashed lines correspond to $\alpha=2$ and $\alpha=1$ respectively.}
\label{alpha}
\end{figure}

The most favored models for type~Ia supernova progenitors invoke a very high 
$\dot{M} \gtrsim 10^{-7} M_{\odot}{\rm yr^{-1}}$, such that the accreted hydrogen 
and helium can burn steadily \cite{cumming}. 
These values of $\dot{M}$ are 
observed in supersoft X-ray sources and symbiotic binaries \cite{cumming,partha}. 
At lower $\dot{M}$s, hydrogen burning is unstable and occurs in flashes, while 
at higher rates an extended envelope is formed \cite{vanden}. 

We now plan to invoke a range of $\dot{M}$s, low to high, in a single model evolution of the white dwarf
and hence consider
\begin{equation}
\dot{M} = \left(\frac{M}{M_{0}}\right)^{\alpha} \times 10^{-9} {\rm yr^{-1}},
\end{equation}
where $\alpha$ is a parameter that determines the functional dependence of $\dot{M}$ on the 
instantaneous mass ($M \geq M_0$) of the white dwarf. 

In Figure 3(b) we show two cases corresponding to $\alpha=1$ and $\alpha=2$, having 
the same final exploding mass, namely $2.33 M_{\odot}$. In the case with $\alpha=1$, $\dot{M}$
varies from $10^{-9}M_{\odot}{\rm yr^{-1}}$ to $1.3 \times 10^{-8}M_{\odot}{\rm yr^{-1}}$. 
The evolution is slow in this case and the supernova explosion occurs in $\sim 5 \times 10^8$ years. 
In the case with $\alpha=2$, $\dot{M}$
varies from $10^{-9}M_{\odot}{\rm yr^{-1}}$ to $1.7 \times 10^{-7}M_{\odot}{\rm yr^{-1}}$. 
The evolution here is more than twice as fast as the previous case 
and the supernova explosion occurs in $\sim 2 \times 10^8$ years.
Hence, $\alpha\ge 2$ is supposed to validate the underlying white dwarfs as 
progenitors of the observed peculiar type Ia supernovae.

\section{Discussion and Conclusions}

Based on the theoretical work on strongly magnetized white dwarfs having 
super-Chandrasekhar masses (DM1, DM2), 
we have discussed possible evolutionary paths that lead from 
magnetized accreting white dwarf binaries to the peculiar type~Ia supernova 
explosions. Multiple evolutionary tracks are possible in the mass-radius 
plane, which covers a mass range $\sim 2-2.6 M_{\odot}$, that 
lies within the observational limits.
We have considered both constant as well as varying $M_\odot$, in order 
to estimate the timescale of occurrence of the supernova events.

Highly magnetized white dwarfs have been discovered 
by SDSS having surface fields in the range $10^5-10^9$ G \cite{schmidt03,vanlandingham05,wick}. 
Note that their central fields are expected to be a few orders of
magnitude higher. 
The properties that we have assumed in our computation are similar to those seen in 
IPs. They have surface magnetic fields ranging $10^5-10^7$ G
and they are high
accretors with $\dot{M} \approx (0.2-4) \times 10^{-9} M_{\odot} {\rm yr^{-1}}$ \cite{warner}.
Cumming (2002) studied the evolution of magnetic fields in accreting white dwarfs
and found that surface magnetic fields are reduced significantly for
$\dot{M}> \dot{M}_{c} \approx (1-5) \times 10^{-10} M_{\odot} {\rm yr^{-1}} $,
due to the advection of the field into the interior of the white dwarf by
the accretion flow.
This presumably increases the central magnetic field \cite{cumming}.
Further, certain dwarf novae (a subclass of CVs) like RU Peg,
are expected to show short and long term variations in $\dot{M}$s, ranging from
as low as $(1-2) \times 10^{-9}M_{\odot}{\rm yr^{-1}}$ to as high as $10^{-7}M_{\odot}{\rm yr^{-1}}$ \cite{partha}.
Hence, it is quite conceivable that IPs with high internal field and low surface field with
an increasing $\dot{M}$ can lead to super-Chandrasekhar white dwarfs causing the 
peculiar supernova Ia explosions.
Moreover, for $\alpha=2$ according to our chosen model, the desired $\dot{M}$ for a type~Ia supernova explosion
to occur is attained starting from an initial, lower rate, typically observed in IPs.
Note that peculiar type Ia supernovae arise, according to our argument, from {\it highly} magnetized 
CVs, which are presumably $10\%$ of 
all magnetized CVs. Hence about 2.5\% of all CVs should lead to super-Chandrasekhar 
objects, which is consistent with the observed rate (1 -- 2\%) of peculiar 
type Ia supernovae \cite{scalzo2012}.

Apart from the advection of the magnetic field 
into the interior caused by accretion,
there are other ways of generating such strong fields.\footnote{However, see the Appendix of DM1, which 
discusses that even a relatively weaker field than that considered here can give rise to the super-Chandrasekhar
masses.} 
Interestingly, recent observations of magnetic Ap and Bp stars indicate
that they have a magnetic flux in the range $10^{26}-10^{27}$ G ${\rm cm^{2}}$, 
which is very similar to that observed 
in magnetized white dwarfs \cite{lila}. This strongly points towards the fact that the magnetic fields 
of highly magnetized white dwarfs are fossil remnants from their 
main-sequence progenitor stars \cite{tout}. Simulations of magnetic field evolution in Ap stars 
strongly support the idea of this field itself being a fossil remnant from the interstellar medium, 
generated during the process of star formation \cite{spruit}. 
Results of these simulations to hold true, the new born star is required to have an initial 
field mainly confined to the core, which can be justified from the flux freezing 
theory during its formation \cite{john}. Moreover, the contraction of a typical interstellar cloud 
of radius $\sim 0.1$ pc, mass $\sim M_{\odot}$ and having a frozen-in magnetic field 
$\sim 3 \times 10^{-6}$ G could give rise to a field $\sim 10^8$ G in the 
resulting star \cite{shapiro} with solar radius. Thus, a star with such a high centrally 
concentrated field 
is a plausible progenitor of the strongly magnetized super-Chandrasekhar white dwarfs 
discussed in this {\it Letter}.
These stars also have convective cores 
and in some special cases a dynamo mechanism might occur to generate a high 
magnetic field therein, leading to a (central) magnetic flux as high as 
$\sim 10^{30}$ G ${\rm cm^{2}}$, 
similar to that of the magnetized white dwarfs mentioned 
in \S 2. The incidence of such main-sequence magnetic stars might be rare, but 
so are the super-Chandrasekhar supernova events.

If the single-degenerate accreting scenario is the correct progenitor model for
the overluminous (peculiar) type Ia supernovae, then
observationally such high magnetic, high mass white dwarfs with small radius 
(also with high $\dot{M}$) could be one of the plausible progenitors of these supernovae. 
However, a detailed investigation needs to be carried out in order to 
understand the physics of explosions of these white dwarfs and if they
indeed produce the light curves similar to that of the peculiar type Ia supernovae.

Now the white dwarfs would be seen as peculiar, super-Chandrasekhar
objects only during a small fraction of their life time, which is $\sim 0.05$ for $\alpha=2$.
Hence, about $\sim 0.12\%$ of the CVs with varying $\dot{M}$ would be likely candidates for
peculiar objects. The varying $\dot{M}$ is further justified by astronomical indications, which
imply higher accretion rates at later times of the evolution of magnetized massive 
white dwarfs \cite{wang-han,toonen,nomoto}. 
We speculate that these white dwarfs
will have X-ray luminosity in between commonly observed accreting white dwarfs and accreting neutron stars
(in the range of $10^{35}$ --$10^{38}$ ergs s$^{-1}$). 
Bright X-ray selected CVs (and other objects) might harbor some of these peculiar super-Chandrasekhar
objects.

\acknowledgments
This work is partly supported by ISRO project Grant No. ISRO/RES/2/367/10-11.
U.D. thanks CSIR, India for financial support. The authors would
like to thank Arnab Rai Choudhuri of IISc for discussion.

\end{document}